\documentclass{article}

\usepackage{arxiv}

\usepackage[utf8]{inputenc} 
\usepackage[T1]{fontenc}    
\usepackage{hyperref}       
\usepackage{url}            
\usepackage{booktabs}       
\usepackage{amsfonts}       
\usepackage{nicefrac}       
\usepackage{microtype}      
\usepackage{graphicx}
\usepackage{natbib}
\usepackage{doi}

\usepackage{tabularx}
\usepackage{stackengine}
\usepackage{rotating}
\newcommand*\rot{\rotatebox{90}}
\usepackage{cleveref}
\usepackage{mathabx}

\title{The Arcanum Mission: Scientific Objectives and Instruments for Neptune, Triton and KBOs}

\date{28th of June, 2021}

\author{Conceptual Exploration Research\thanks{\href{https://orcid.org/0000-0002-4071-5727}{\includegraphics[scale=0.06]{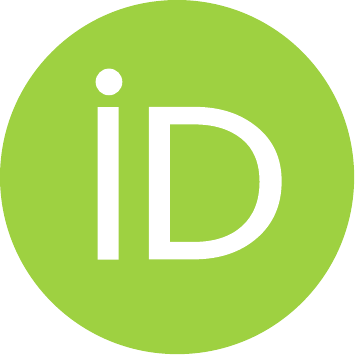}\hspace{1mm}James E.~McKevitt}, Christina Bornberg, Tom Dixon, Louis Ayin-Walsh, Jonathan Parkinson-Swift, James Morgan, Shayne Beegadhur, Franco Criscola, Carina Heinreichsberger, Bharath Simha Reddy Pappula, Sophie Bulla, Kuren M. Patel, Aryan Laad, Ethan Forder, Jaspreet Singh, Oisín Moore, Madalin Foghis, Paul Wedde, Thomas Mcdougall, Jack Kent, Utkarsh Raj} \\
\texttt{hello@conexresearch.com}}

\renewcommand{\headeright}{Conex Research}
\renewcommand{\shorttitle}{Arcanum: Scientific Objectives and Instruments}

\hypersetup{
pdftitle={The Arcanum Mission: Scientific Objectives and Instruments for Neptune, Triton and KBOs},
pdfsubject={astro-ph.IM},
pdfauthor={Conceptual Exploration Research},
pdfkeywords={Neptune, Triton, KBO, Penetrator, Telescope},
}

\begin{document}
\maketitle

\begin{abstract}
The Arcanum mission is a proposed L-class spacecraft that highlights the revolutionary approach which can now be taken to future space mission design. Using the case of the SpaceX Starship vehicle and in particular the high mass and volume characteristics of this launcher, the feasible large size of future missions, even with high delta-V transfer requirements, are analysed. A demonstrator vehicle, designed to support a large and capable science platform with multiple components, is detailed, clearly showing the range and depth of science goals that will be answerable thanks to the current revolution in super heavy-lift launch vehicles. 
\end{abstract}


\section{Introduction}
As stated in the Planetary Science Decadal Survey (2013-2022) \citep{CommitteeonthePlanetaryScienceDecadalSurveySpaceStudiesBoard2011Vision2013-2022}, a Neptune System orbiter and probe is of high scientific interest but lacking in supporting technical know-how. Therefore, such an undertaking is currently labelled as a \lq{}deferred high-priority mission\rq{}, with a focus required around optimising transfer trajectories, lowering the cost of delivery, further understanding aerocapture abilities, providing suitable power and propulsion, high-performance telecommunications and thermal protection. Notably, a flyby is not advised given its costly nature with relatively little science return. As highlighted by the Neptune Odyssey mission concept \citep{Cohen2020NeptuneSystem}, when comparing missions to both Neptune and Uranus, only the former offers both an ice giant environment and a captured Kuiper Belt Object (KBO).\par
Arcanum is classified as an L-class mission, making it a flagship endeavour. It is common for such missions to see a launch cadence of approximately one launch per decade due to key constraining factors such as mission pricing, scale and longevity. The high tonnage of an L-class mission such as Arcanum means only a super heavy-lift launch vehicle is capable of delivery of the mission onto an interplanetary trajectory.\par
To support the Arcanum mission, three main launch vehicles were considered:
\begin{itemize}
    \item NASA's Space Launch System (SLS) Block 2 Cargo
    \item Space Exploration Technologies Corp's Starship
    \item Blue Origin Federation's New Glenn
\end{itemize}
This list was constrained by a number of factors, including a forecast future geopolitical environment.\par
Out of these three upcoming launchers, SpaceX Starship has been selected as the most viable option. This is due to the vehicle's intended rapid reusability and low operational launch costs. SpaceX claims the Starship system could cost as little as \$2 million per flight, severely undercutting preliminary costings of the SLS Block 2 Cargo at \$2 billion per flight. The Starship programme has also received a monetary injection by the U.S.A.F. \citep{USDepartmentofDefenseAirForce2021FiscalVol-1} and the NASA Human Landing System (HLS) Program. This provides confidence in Starship’s operational capabilities by the early 2030s. Additionally, the payload fairing volume makes Starship a compelling option for deploying Arcanum, a particularly large-volume spacecraft.
\section{Spacecraft Architecture}
\begin{figure*}
    \centering
    \includegraphics[width=\columnwidth]{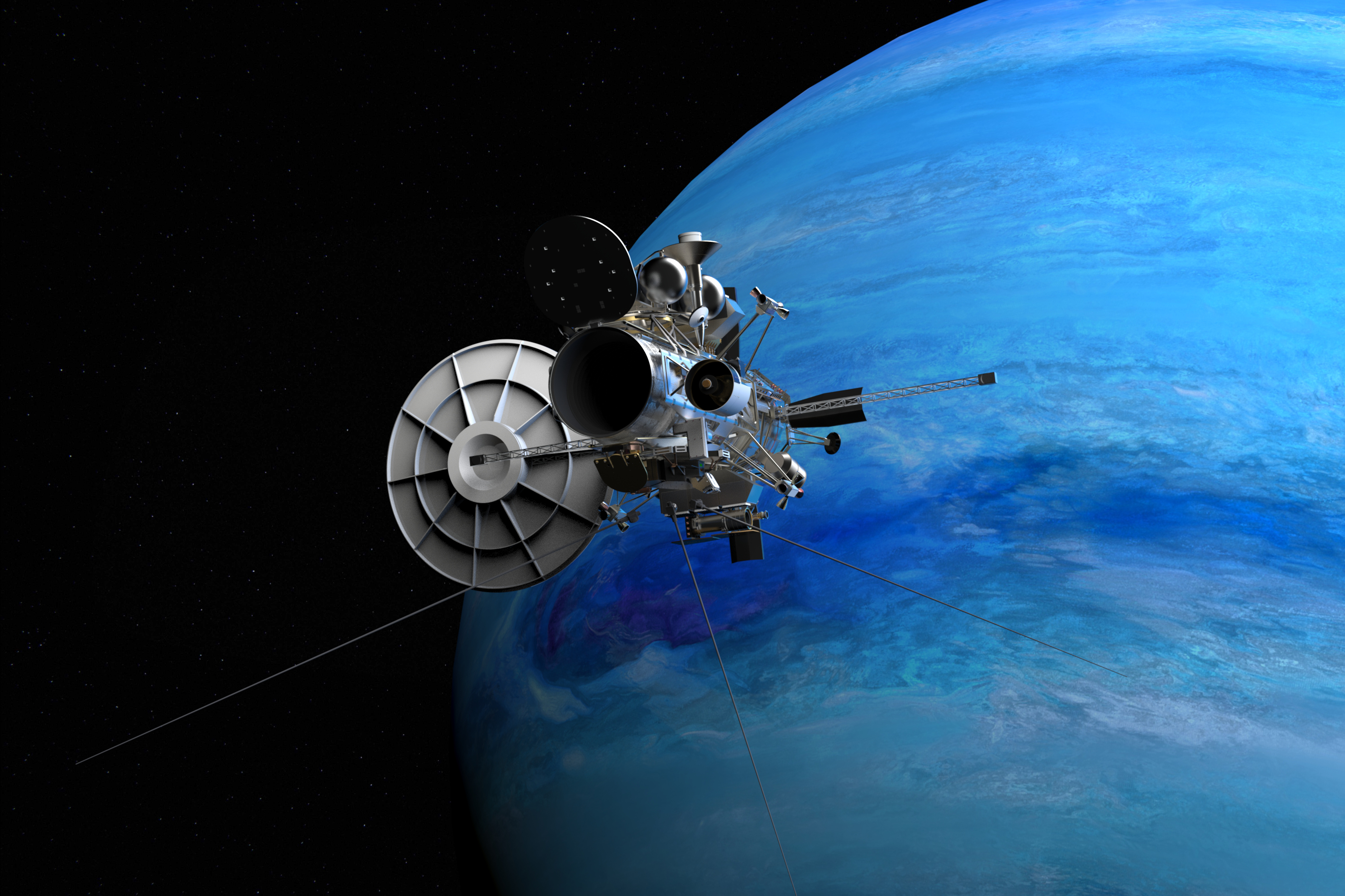}
    \caption{The Arcanum mission’s Somerville-Bingham spacecraft in orbit around Neptune.}
    \label{fig:cover}
\end{figure*}
\subsection{Somerville}
The Arcanum mission architecture comprises a number of key components, the primary of these being the Somerville Orbiter. This is a large spacecraft bus designed around a central, 1-metre diameter telescope, capable of supporting a number of constituent spacecraft fulfilling the wide-ranging science goals which classify this spacecraft as one performing multiple roles. The operation of this spacecraft, details of which were recently published by \cite{McKevitt2021}, centres around its placement in a highly eccentric Neptunian orbit. From this location, it will be capable of addressing science goals on objects in the Neptunian System, those surrounding Kuiper Belt Objects, and will continue to develop our understanding of exoplanetary systems. A previously undisclosed ability relating to these exoplanetary systems is detailed below. However, a comprehensive description of the wide-ranging targets this spacecraft will be able to address is described in the aforementioned paper.\par
The power systems of the spacecraft have also been previously described. However, the required power generation capability of the RTGs used dictates a highly abnormal number of Cassini-sized RTG units. The total required by the entire Arcanum mission is approximately 26, a number providing a reasonable margin, allowing the addition of further instruments or spacecraft capabilities. This number has been calculated using the available data on the Americium-241 RTG units currently under development in Europe.\par
\subsection{Bingham}
The soft-landing Bingham probe, a science platform that will descend to the surface of Triton, a key moon of Neptune, is the other key component of this mission, joining with Somerville to form the complete Somerville-Bingham spacecraft pairing. This coupled unit will launch and transit to Neptune together, with Somerville providing power, thermal and communications support for the lander on the journey. Once placed in a highly-eccentric Neptunian orbit, Bingham will be released into a trajectory ready for landing on Triton. This course will be created by the orbiter, minimising the delta-V, and therefore propellant, required for the landing. This aims to maximise the number of scientific instruments capable of operating on the spacecraft, and therefore the science return from one of the most intriguing destinations in our Solar System.\par
Of all the worlds where a soft landing has been attempted, Earth’s Moon is most similar to Triton. Perhaps not when it comes to composition and the ambient temperature, but these two factors are more important for long-term operations. For the landing two other factors are key: the lack of an atmosphere and the gravity. Mars is dissimilar to Triton due to its atmosphere, with an average surface pressure of $0.636kPa$, thin but not thin enough to be comparable as Triton’s atmosphere is around 385 times less dense on average, and possesses much higher surface gravity at $3.72m/s^2$. Asteroids on the other hand, while also devoid of an atmosphere, have almost no gravity at all. For this reason, the landing methods of lunar probes were considered.\par
The most common choice is propulsive landing via rockets. Most descent vehicles, from the Viking landers to the skycranes of Curiosity and Perseverance, have used storable monopropellant engines, with the exception of Starship and Blue Moon. These two conceptual landers use cryogenic propellant.\par
An alternative was to employ the mechanism used in the first lunar landers, specifically Luna 9, which since its landing in 1966 has been rarely utilized again. Luna 9’s landing used rockets, with the Luna unit being divided into two parts, the lander and the payload. The lander used Doppler and retro rockets for descent and control and after the landing was completed, the payload would hop onto the ground and open its outer protective shell. This was done because at that time not much was known about the Moon, and this method was deemed to decrease the risk of failure. A similar methodology could be implemented given similar uncertainties in Triton’s environment.\par
It is likely that a thruster-based system will be utilised for all or part of the terminal descent of the Bingham Lander. With the desire to maximise the lander mass available for instruments whilst minimising the total wet mass, a staged combustion engine cycle or similar would not be suitable. Looking to minimise complexity and simplify the construction of the spacecraft, while also limiting the potential for mechanical failure to arise during the coast phase of the mission, a pressure-fed or expander cycle system would appear most attractive. The expander cycle offers a high specific impulse but is hampered by the need for cryogenic fuel, the management of which would add significant operational complexity to the spacecraft: to prevent boil-off, the fuel must be shielded from even minimal temperature fluctuations. To date, even advanced storage concepts would suit mission durations of 60-90 days, which is far too short for this mission and would require substantial mass gains \citep{Doherty2013CryogenicTDM}. As such, Bingham employs a pressure-fed system. This offers a high specific impulse and a comparatively low associated \lq{}cost\rq{} in dry mass, due to the lack of required plumbing. The need to optimise propellant pressure against added fuel tank mass means there are practical limits on the combustion chamber pressure. However, the low gravity of Triton means that this is acceptable as the minimum required thrust for safe deceleration remains low.\par
A precedent established by the unmanned interplanetary landers to date is to utilise hydrazine-fuelled engines. The wealth of operational time and experience (TRL 9) of similar systems makes them an attractive choice. Hydrazine and its derivatives can be utilised in both monopropellant and bipropellant systems. The monopropellant option has more precedents when it comes to landers such as Phoenix and Schiaparelli, as well as the Curiosity and Perseverance skycranes, simplifying the ignition process and reducing the likelihood of a hard start which would be mission-fatal at Triton. Monopropellant engines also add additional flexibility in the form of being reignitable and offering a high degree of throttleability, making them ideally suited for descent engines. The most suitable and means-tested fuel mixture consists of liquid hydrazine passed over a granulated iridium-coated alumina catalyst prior to combustion \citep{Harden1965ThermodynamicHydrazine}. The resulting decomposition is highly exothermic and can be harnessed to produce up to 400N of thrust at over 200s ISP in existing motors. Other alternative fuels are under consideration in the event that ESA moves to limit the use of hydrazine and other hazardous fuels. A strong cleaner alternative with similar, if not improved, propulsive properties would be hydroxylammonium nitrate \citep{Spores2014GPIMSystem}.\par
Airbags are considered as a means of mitigating the forces experienced during landing on the spacecraft. They can allow reduced fuel loads and can make the spacecraft more durable when it comes to terrain conditions and orientation at impact. A principal concern was the effect of thruster plumes on the chemical structure of the landing site: there are concerns that this would invalidate some experiments. There is also the risk that melting of the local regolith, which is mostly comprised of frozen nitrogen and water, due to engine plumes, could cause the lander to become unstable \citep{Holler2016OnLatitudes}. This could be exacerbated if the landing site is already structurally weak. An airbag system would offer a solution to these difficulties.\par
Heritage airbag landing systems include early Soviet lunar landers, where airbags were used after landing to surround and cushion an ejected payload module, and more recently the NASA Mars Exploration Rover Mission operated a more advanced system of lobed Vectran airbags arranged in a tetrahedral structure around the rovers due to engine plumes, could cause the lander to become unstable \citep{Crisp2003MarsMission}. This study will also consider airbags used purely to absorb the impact shock of an already-righted spacecraft, in the same manner as the Boeing Starliner. Whilst airbags do help solve the problems of chemical contamination of the landing site, they also lack the precision landing ability of engines, and are still dependent on a partially powered descent to ensure the spacecraft is travelling at acceptable speeds at impact. The airbags must also deploy rapidly through the use of heavy gas generators and allow the spacecraft to right itself if it lands in an incorrect orientation, then retract to leave the lander set on a rigid structure, something essential for operation of the seismometers. With this in mind, a hybrid of the Starliner method and the engine system described above would appear preferable, if airbags are to be used.\par
Whilst airbags are still under consideration, a less operationally complex alternative would be to opt for a full-powered descent and mechanical landing legs. The thrusters could be deactivated at an altitude that would cause minimal surface contamination or melting, and the legs absorb the remaining impact shock. An optimal design based on past research would involve telescopic legs with aluminium honeycomb crush core inserts and partially deformable domed footpads. The honeycomb inserts deform to absorb impact energy; an insert in the primary strut is more efficient as all forces are applied along the crushpad’s main axis, which is beneficial as lateral compression drastically reduces the material’s absorption efficiency. Telescopic legs are responsive to uneven ground conditions, improving the versatility of the lander, and the domed footpads exhibit greater strength than a flat pad of the same mass and can deform in response to uneven regolith, such as in the presence of rocks. This being the case, along with the need to have a propulsion system regardless of whether or not airbags are used, a system of landing legs like those described would appear to be a strong contender \citep{Sahinoz2012LandingLander}.
\subsection{Penetrators}
\begin{figure}
    \centering
    \includegraphics[width=.4\columnwidth]{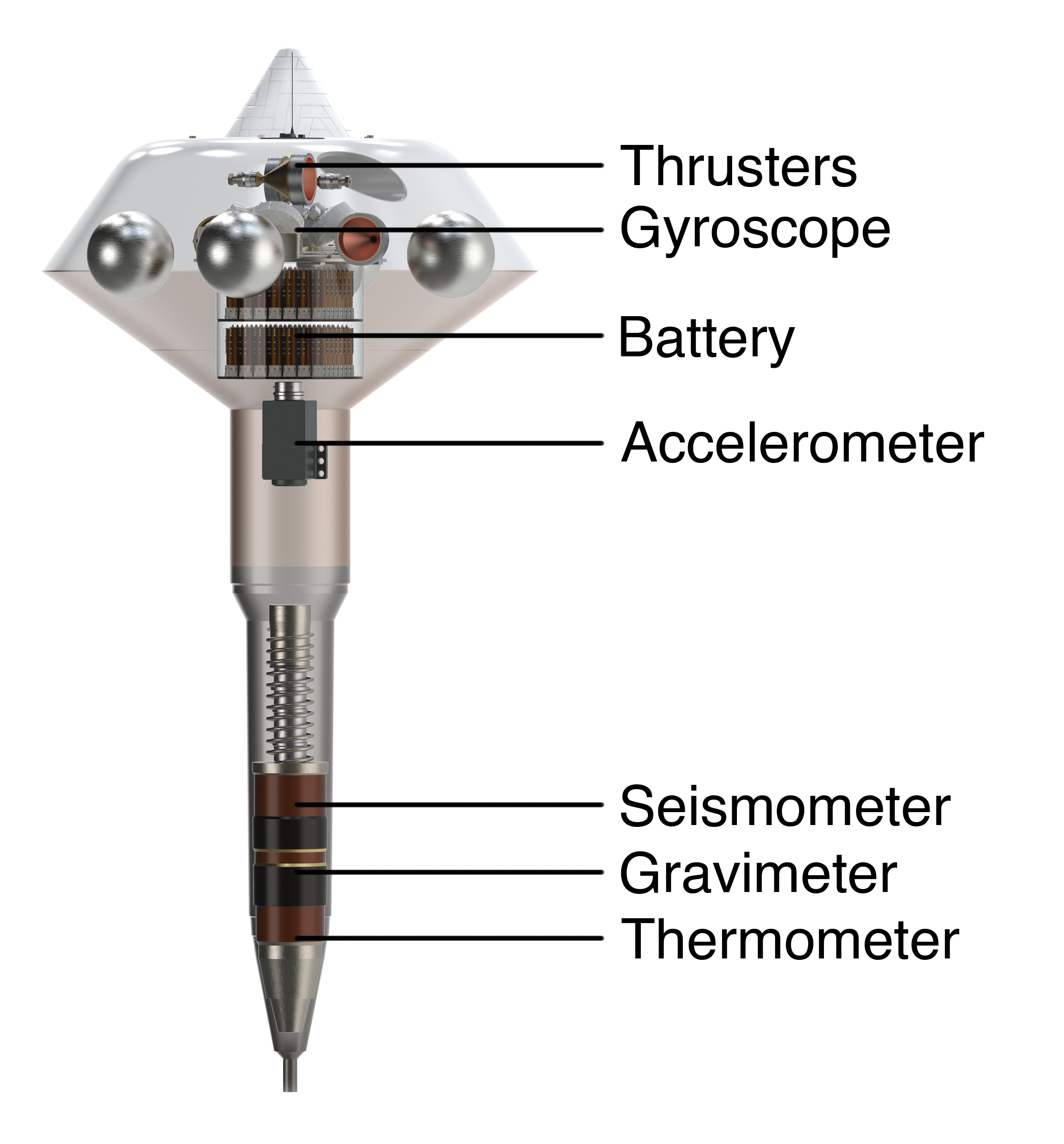}
    \caption{A penetrator used on the Arcanum mission, with key components highlighted.}
    \label{fig:penetrator_diagram}
\end{figure}
\begin{figure}
    \centering
    \includegraphics[width=.4\columnwidth]{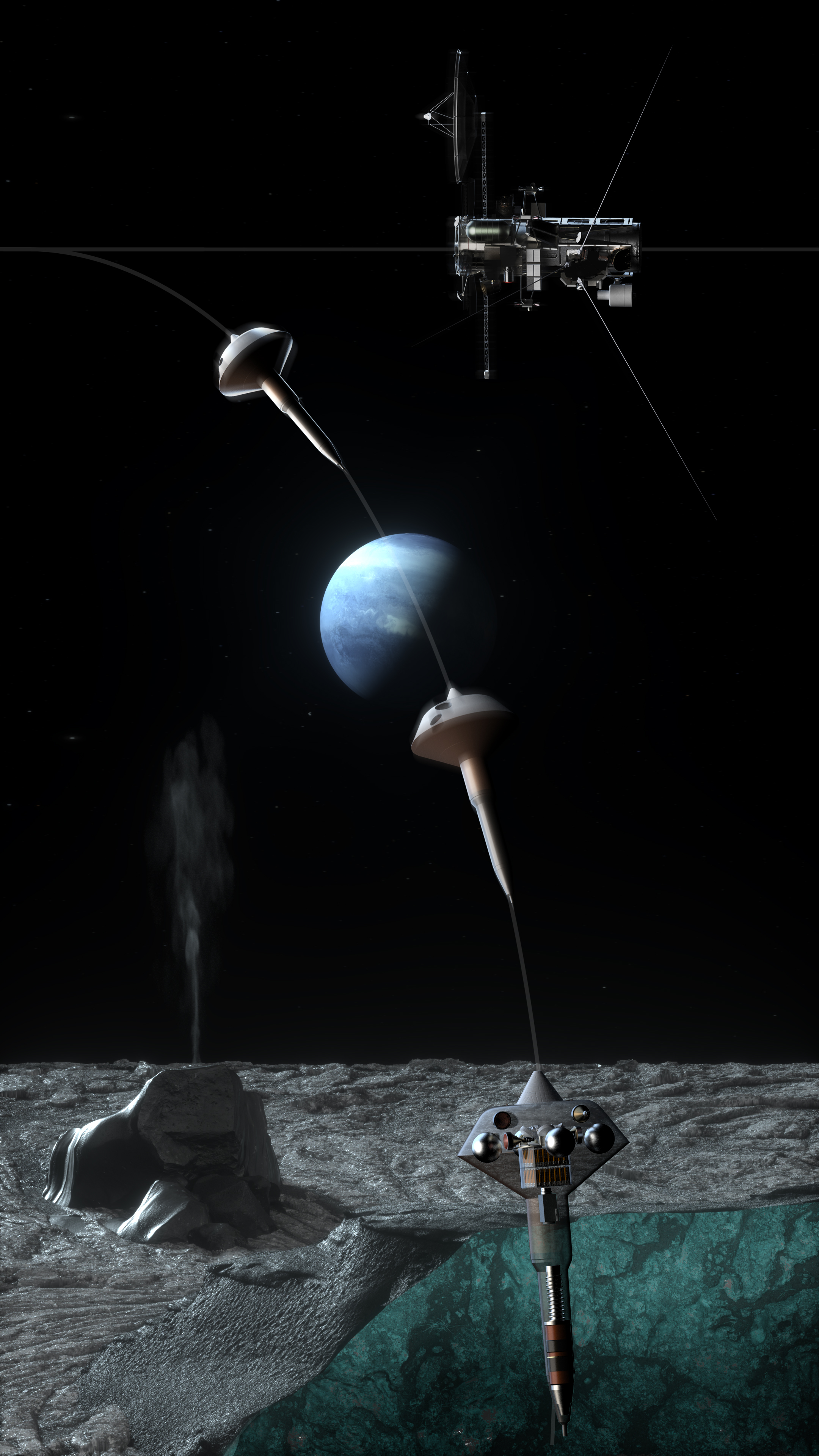}
    \caption{A graphic illustrating the penetrator used on the Arcanum mission following deployment from the Somerville Orbiter.}
    \label{fig:penetrator_deployed}
\end{figure}
As previously stated, Arcanum consists of an orbital deep space telescope and landing probe, named Somerville and Bingham respectively, the former of which deploys two surface penetrating probes. Due to the low density of Triton’s atmosphere, the use of multiple penetrators in conjunction with a small planetary lander was selected, allowing for a greater operational area for the undertaking of scientific measurements. This coverage can be further increased by the inclusion of additional penetrators, albeit at the expense of increased mass, or the adoption of micro penetrators, increasing the number supported onboard of Somerville for the equivalent mass of larger less numerous penetrators. The result of this is a network of numerous micro penetrators located at various geographically and/or environmentally diverse locations, thereby increasing the potential scientific opportunities \citep{Ahrens2021SmallScience} whilst providing increased mission redundancy. This publication will introduce and focus on the use of larger penetrators onboard Somerville, however further studies will be conducted into the use of micro penetrators, something to be detailed in further work.\par
A review of previous and current penetrator designs was undertaken to determine the design requirements needed for the Arcanum mission in conjunction with the development of a feasible penetrator concept. The previously discussed Mars96 mission was used as the basis of the penetrator concept due to the project achieving a fully developed design, unfortunately however failing to reach Mars due to a launch vehicle failure \citep{Lorenz2011PlanetaryFuture}.\par
The initial Arcanum penetrator concept, like the Mars-96 design, would feature a fore and aft body separated by a conical flared guard towards the aft section to ensure the communication systems are exposed to the planetary surface, as illustrated in \ref{fig:penetrator_deployed}.\par
\begin{figure}
    \centering
    \includegraphics[width=.8\columnwidth]{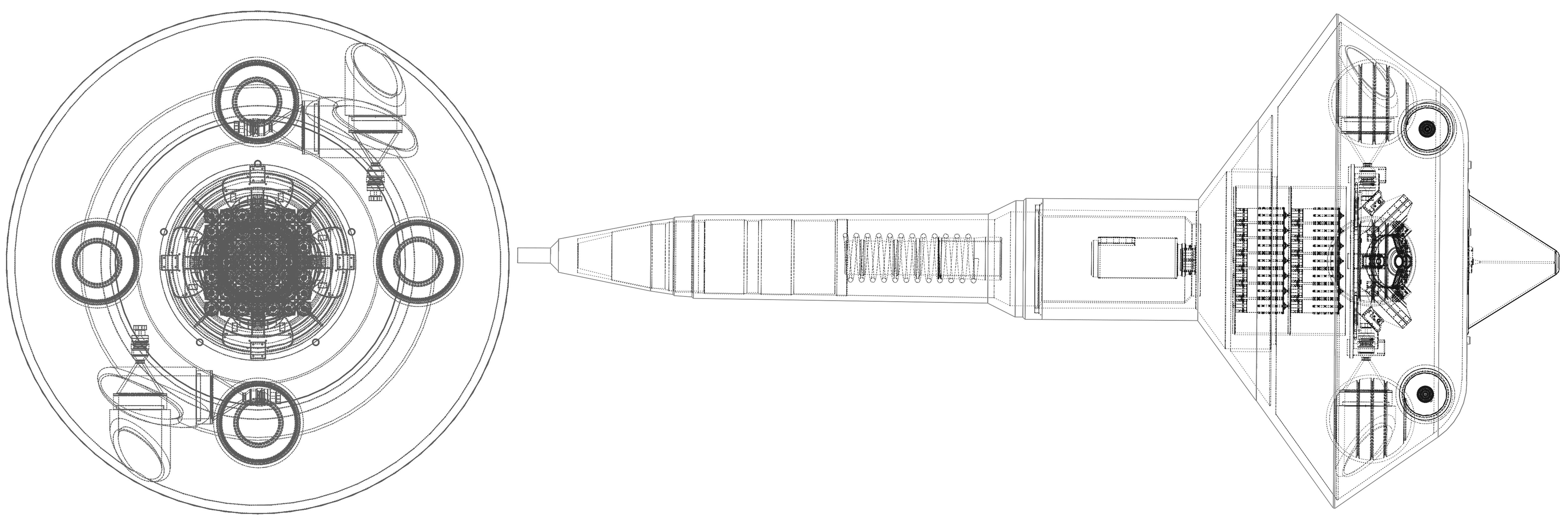}
    \caption{A penetrator used on the Arcanum mission.}
    \label{fig:penetrator_wireframe}
\end{figure}
\begin{figure}
    \centering
    \includegraphics[width=.6\columnwidth]{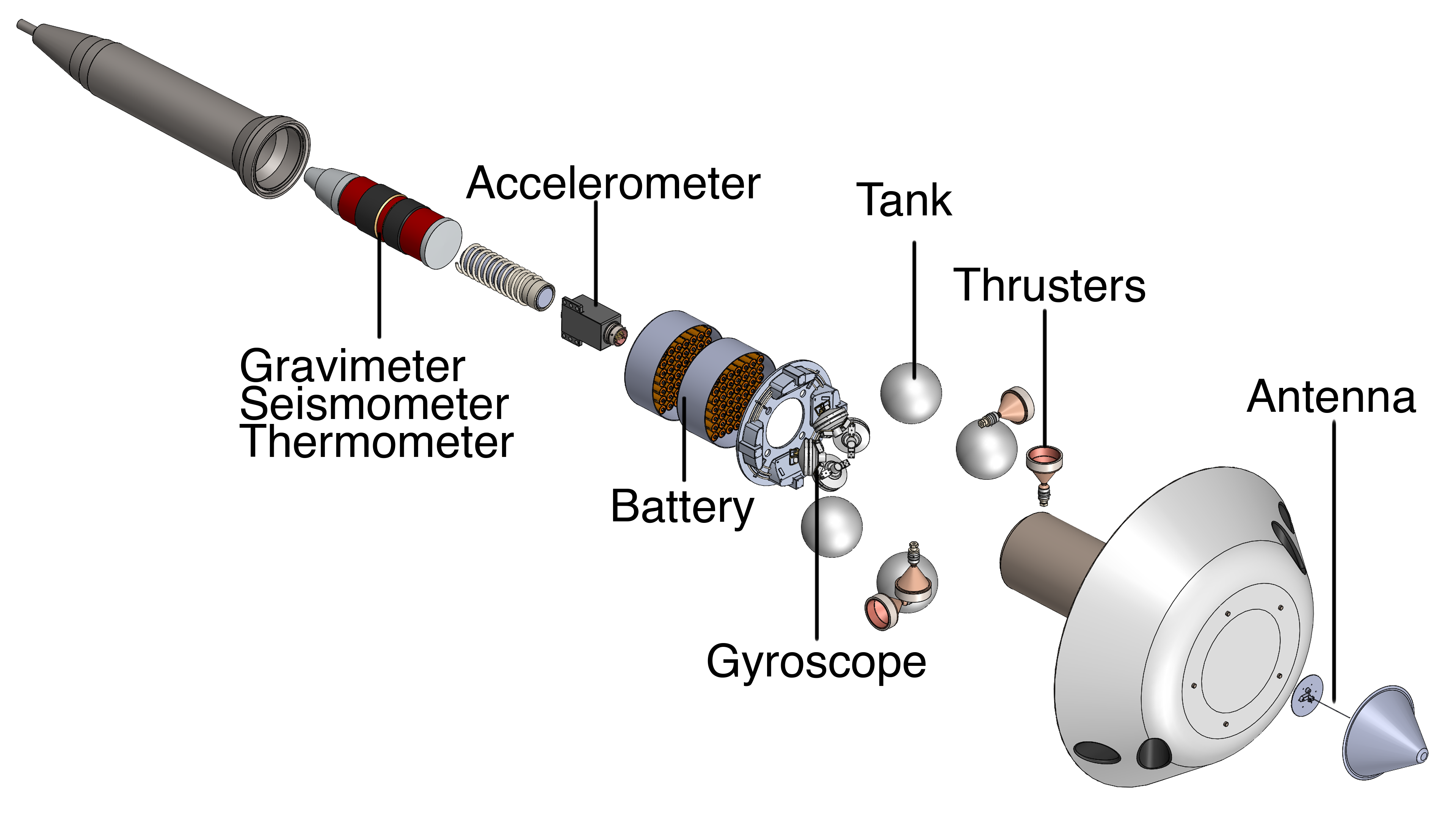}
    \caption{Exploded view of the Triton penetrator.}
    \label{fig:penetrator_exploded}
\end{figure}

However, unlike the Mars-96 design, an inflatable decelerator, known as a ballute, would not be employed within the Arcanum design due to the lack of an atmosphere at Triton \citep{Lorenz2011PlanetaryFuture}. The initial concept calls for the use of RCS thrusters for the control of orientation during the descent phase, whilst an onboard descent camera mounted towards the aft section will provide images of the Triton surface and the penetrator forebody. Once impact occurs, the onboard accelerometer will be used to measure mechanical properties of Triton’s surface by the response of the projectile, such as the deceleration, entry oscillations, and penetration depth. It was established a penetration depth of approximately one metre would be suitable for the recording of these measurements. Therefore, the initial design concept would have a forebody and total length of approximately 1m and 1.4m respectively. The forebody will have a feasible diameter to accommodate the internal components whilst the protruding tip has a diameter of 0.022m. Further sizing of the penetrator will be determined as the design process moves from conceptual design through to the embodiment and subsequently detail design phases. To supply power to components throughout the operating period, the penetrators will utilise non-rechargeable lithium thionyl chloride batteries. These are selected for their high energy density (over 700Wh/kg), long lifespan, high rigid body shock tolerance, consistent voltage throughout the discharge period and consistent performance at low temperatures. The design is modelled off the batteries used in the NASA Deep Space-2 Mars impactor mission \citep{Smrekar1998DeepBeyond}. The internal components and their subsequent layout within the penetrator are illustrated in \Cref{fig:penetrator_diagram,fig:penetrator_wireframe,fig:penetrator_exploded}.\par
The outlined concept above will undergo multiple design iterations evaluating its feasibility, future additional requirements and design alterations, assessing the optimum dimensional and material properties to be employed on the penetrator design. Further evaluations will be conducted by performing high-fidelity contact simulations between the penetrator and the planetary surface using LS-DYNA, evaluating the effects on both during impact. A high-fidelity structural analysis will be performed using finite element analysis (FEA) software, evaluating the stress incurred by the penetrator during impact thus aiding in material selection and the implementation of structural reinforcements if required. Details of these and a further in-depth outline of the proposed penetrator design concept will be reserved for future publications \citep{Ahrens2021SmallScience}.
\section{Science Overview}
The science goals of the Arcanum mission evolve around the Neptunian system; including Triton, the Kuiper Belt and other objectives that can be answered with the telescope, part of the Somerville Orbiter. There are a number of pieces of literature proposing science goals for this system, particularly Neptune and Triton \citep{CommitteeonthePlanetaryScienceDecadalSurveySpaceStudiesBoard2011Vision2013-2022,Masters2014NeptunePuzzle,Helled2020TheQuestions} which have been used for inspiration. Furthermore, details of the type and method of Neptunian science can be found in previous work on this study \citep{McKevitt2021}.\par
The instruments for the Somerville Orbiter (S) and its Telescope (T), the Bingham Lander (B) and the Penetrators (P) are presented in two traceability matrixes: \Cref{tab:tracability_neptunetriton,tab:traceability_remote}.\par
\begin{table*}
\centering
\caption{Traceability Matrix \lq{}Neptunian System: Neptune (N), Triton (T)\rq{} including instruments: seismometer, gravimeter, thermometer, accelerometer, camera, spectrometer, and magnetometer and magnetospheric imager (Mag). Part of 2 Penetrators (P), Bingham (B) and Somerville (S). 1: Primary, 2: Secondary.}
\label{tab:tracability_neptunetriton}
\begin{tabularx}{\textwidth}{|X|l|l|l|l|l|l|l|l|l|l|}
    \hline
    &
    \rot{\textbf{Seismometer (B\&P)}} & 
    \rot{\textbf{Gravimeter (P)}} & 
    \rot{\textbf{Thermometer (B\&P)}} & 
    \rot{\textbf{Accelerometer (B\&P)}} &
    \rot{\textbf{Cameras (Bx2\&P)}} & 
    \rot{\textbf{Spectrometers (B)}} & 
    \rot{\textbf{Mag (S)}} &
    \rot{\textbf{Spectrometers (S)}} & 
    \rot{\textbf{Cameras (Sx2)}}
    \\\hline
    \textbf{N-1. Origin}: Where and how did Neptune form and what Nobel gases are present? Has Neptune collided with another body? How did the rings form?
    &-&-&-&-&-&-&1&1&1
    \\\hline 
    \textbf{T-1. Origin:} When was Triton captured and how has it evolved?
    &1&1&2&1&2&1&1&1&2
    \\\hline 
    \textbf{N-2. Interior and subsurface:} 
    Why is Neptune's measured intrinsic heat flux so high? Why do Uranus and Neptune have heavier elements than Jupiter and Saturn? What is the origin and abundance of light elements in the deep interior?
    &-&-&-&-&-&-&2&1&2
    \\\hline 
    \textbf{T-2. Interior, surface and geological activity:} 
    Is there an ice cap on the North Pole? Is Triton (still) geologically active? Is the interior of Triton warm enough to generate a convective current in the layers? What is the age and what are the geological processes that shape the surface of Triton/KBOs. What is the hardness of the surface?
    &1&1&2&1&1&1&2&1&1
    \\\hline
    \textbf{N-3. Atmosphere and Weather:}
     What are atmospheric structure and cloud properties? What is the origin and abundance of ices in the atmosphere and outer envelope?
    &-&-&-&-&-&-&1&1&1
    \\\hline
    \textbf{T-3. Ionosphere and Seasons:} Why is Triton's ionosphere so dense? How do seasons vary on Triton? How is geyser activity related to the seasons, Triton experiences?
    &-&-&-&-&1&1&1&2&1
    \\\hline
    \textbf{N-4. Magnetic field and offset:} Are Neptune's auroral emissions similar to the ones on Earth, Jupiter and Saturn? What is the origin and structure of Neptune's offset magnetic field? 
    &-&-&-&-&-&-&1&-&-
    \\\hline
    \textbf{N-5. Moons and Rings:} How did the inner satellites form? What is the reason for the \lq{}Dance of Avoidance\rq{} of Naiad and Thalassa?
    &-&-&-&-&-&-&1&1&1
    \\\hline
\end{tabularx}
\end{table*}
\begin{table*}
\centering
\caption{Traceability Matrix \lq{}Remote (R)\rq{} including instruments: camera, spectrometer, photometer, magnetospheric imager (Mag) and dust detector. Part of Telescope (T) and Somerville (S). 1: Primary, 2: Secondary.}
\label{tab:traceability_remote}
\begin{tabularx}{\textwidth}{|X|l|l|l|l|l|}
    \hline
    & \rot{\textbf{Camera (T)}} & 
    \rot{\textbf{Spectrometer (T)}} & 
    \rot{\textbf{Photometer (T)}} & 
    \rot{\textbf{Mag (S)}} & 
    \rot{\textbf{Dust detector (S)}}
    \\\hline
    \textbf{R-1. Kuiper Belt Formation:} What were initial stages, conditions and processes of Solar System formation? What is the reason for highly eccentric orbits of Kuiper Belt Objects? Is there evidence of a ninth Planet? How accurate are KBO positions? What is the composition of KBOs?
    &1&1&1&-&-
    \\\hline 
    \textbf{R-2. Constrain Exoplanets}: How do exoplanets influence a star with micro-lensing events? How accurate are current star positions from Earth parallaxes compared to a Neptune-Earth parallax?
    &1&1&1&-&-
    \\\hline 
    \textbf{R-3. Cosmic Dust:} How accurate are zodiacal dust models? To what degree does the zodiacal light region interfere with measurements of extragalactic background light (EBL)?
    &1&1&1&1&1
    \\\hline 
    \textbf{R-4. Instrument Verification:} How accurate are spectrometers and cameras based on and around Earth observing exoplanets? How different are current Earth models when compared to remote sensing the Earth from Neptune?
    &1&1&1&-&-
    \\\hline
\end{tabularx}
\end{table*}
By utilising a highly eccentric orbit around Neptune, observation of the entire Neptunian System is possible allowing the study of all of its moons, including the 7th largest in the solar system, Triton.\par
\subsection{Triton}
Triton is unique in that it orbits Neptune in an opposing direction to its rotation, a phenomenon called a retrograde orbit. It is believed that the moon is a former KBO captured by Neptune, and so a more detailed and dedicated study of the moon and all its characteristics is required to solve the mystery of its origins.\par
The landing location of the Bingham probe could be constrained by several factors. The most critical for meeting the needs of primary and secondary objectives are identified as L1, L2 and L3.\par
\begin{table*}[!htpb]
\centering
\caption{Landing site (L) requirements for the Bingham Lander and Penetrators.}
\label{tab:landing_sites}
\begin{tabularx}{\textwidth}{|l|X|X|X|}
    \hline
    \textbf{ID} & \textbf{Requirement} & \textbf{Experiment} & \textbf{Justification}
    \\\hline
    \textbf{L-1} & Near an active geyser & Analysis of geyser ejecta composition & To collect data as evidence that Triton harbours the building blocks for life
    \\\hline
    \textbf{L-2} & Western hemisphere site & Analysis of ground composition & To collect better data about the “Cantaloupe Terrain” 
    \\\hline
    \textbf{L-3} & Latitude near the subsolar point & Spectral analysis of the ground & To collect data to improve theories on cryovolcanism
    \\\hline
\end{tabularx}
\end{table*}
\subsubsection{L1 - Near an active geyser}
Triton is comprised of mainly nitrogen, water, carbon dioxide and carbon monoxide \citep{McKinnon2014Triton}. The most important of these molecules is water as it primarily exists as ice. However, there is thought to be enough rocky material in Triton’s core to mean radioactive decay can sustain a liquid subsurface ocean \citep{Nimmo2015PoweringGeology}. Coupled with tidal heating caused by Triton’s strong obliquity, it is possible that such a subsurface ocean could harbour some form of life.\par
This could be confirmed by ultrasonic analysis of the crust structure to identify the relative densities and to characterise the crust composition. In conjunction with this, \lq{}capture and characterise\rq{} experiments on ejecta from geysers would further our understanding of their chemical makeup. Black material ejected from Triton has been observed by Voyager 2 and it is suspected that this contains organic matter. Experiments on this ejecta would go a long way to support the theory that the building blocks for some form of life are present on Triton.\par
\subsubsection{L2 - Western hemisphere site}
The advantage of sending a probe to the surface of Triton is that greater analysis of certain geological features can be conducted to reveal information that a standard flyby mission cannot. Most notably would be the \lq{}Cantaloupe Terrain\rq{} located in the western hemisphere of Triton. It is considered to be the oldest terrain on Triton and this sort of surface topology is unique in the Solar System \citep{McKinnon2014Triton}. The results would provide greater evidence to allow scientists to decide whether diapirism or cryovolcanic collapse is the cause for this topology \citep{Smith1989VoyagerResults}.
\subsubsection{L3 - Latitude near the subsolar point}
A leading theory for cryovolcanism on Triton is the \lq{}solid greenhouse effect\rq{} \citep{Smith1989VoyagerResults}. The surface consists of a translucent frozen nitrogen layer on top of a darker substrate. This allows for solar radiation to pass through the surface nitrogen layer and vaporise subsurface nitrogen resulting in an increased pressure below the surface, leading to eruptions. It is estimated that a 4K temperature difference between the surface and subsurface layer is sufficient to generate the pressure to cause the 8km tall plumes as observed by Voyager 2 \citep{Soderblom1990TritonsCharacterization}. A possible measurement technique would be to use a modified version of the Surface Science Package (SSP) as used on the Huygens probe. This would allow for such an experiment to take place.\par
A summary of the three sites and their respective benefits is shown in \ref{tab:landing_sites}.\par
L1 and L3 go hand-in-hand as the majority of geysers are located at 50\textdegree-57{\textdegree} South. This is near the subsolar point ensuring that both constraints are met. However, interpretations of Voyager 2 data indicate that there are virtually no geysers active near the \lq{}Cantaloupe Terrain\rq{}, with there being only one geyser located at the border between the south polar ice cap and this region. Despite this, Triton is a geologically active moon, meaning that the features observed by Voyager 2 will likely have changed by the time of the Arcanum mission. Additionally, the omittance of the Trident mission from NASA’s Discovery Program means that newer information about the location of these geysers will not be available.\par
This lack of information means that additional landing probes will be required to ensure the requirements of L1, L2 and L3 are met. To solve this, two penetrators (one to land in the Western hemisphere and one to land near the subsolar point) will be sent along with the Bingham probe. Prior to the jettison and entry of these probes, a period of the mission will be dedicated to observing and identifying suitable locations for Bingham and the penetrators. Following this, Bingham will be the first probe to land followed by the penetrators to conduct a large-scale seismic experiment. This methodology ensures all requirements of Triton surface experiments are met.
\subsection{Kuiper Belt Objects}
Past the orbit of Neptune lies the Kuiper Belt, a region of space filled with icy objects that orbit the Sun in diverse orbits, much like the asteroid belt. These objects are far too small and dim to be observed by Earth-based telescopes, and so a telescope orbiting Neptune would be invaluable.
\subsection{The Planet Nine Hypothesis}
Previous work has constrained the location of Planet Nine, admittedly within a wide envelope, to the following \citep{Batygin2019TheHypothesis,Batygin2016EvidenceSystem}:
\begin{itemize}
    \item Period: 10,000 - 20,000 years
    \item Semi-major Axis: 400 - 800 AU
    \item Eccentricity: 0.2-0.5
    \item Inclination: 15-25
    \item Argument of Pericenter: 150\textdegree
    \item Mean Anomaly: 0.5
    \item Mass: 5-10 $M_\Earth$
\end{itemize}
Researchers have identified an exoplanet HD 106906b, roughly 11 times the mass of Jupiter, situated some 336 lightyears from Earth, in a 15,000 year-long highly eccentric orbit around a 15 million-year-old binary star system. The planet is separated at a distance of 737 AU from the binary and is still glowing from internal heat. The presence of this exoplanet shows that Planet Nine-like architectures can form early in the evolution of planetary systems. With the help of Hubble imagery and ESA’s Gaia Observatory, the researchers have been able to constrain the measurements confirming that the exoplanet is indeed bound to the binary star system with its orbit inclined at 36 to 44 degrees to the protoplanetary disk. This observation gives credence to the theoretically proposed Planet Nine in the outer reaches of the Kuiper Belt, and provides an interesting secondary objective for the Arcanum mission \citep{Nguyen2020FirstOrbit}.
\subsection{Parallax and micro-lensing observations}
When one star passes another background star within one arcsecond, its gravity acts as a lens, interacting with light in such a way as to cause the background star to appear in a different position. The image of the background star is distorted and depending on the alignment of the two stars, it is either distributed around the lensing star as a ring, known as an Einstein ring, or divided into multiple distorted images close to the Einstein radius. Due to this distortion, the received signal is protracted across a wider area, thus creating a larger brightness signature.\par
If the lensing star is accompanied by an exoplanet, a second peak in the light curve will appear due to the gravitational influence of the exoplanet. How much of an impact the exoplanet has depends on a number of factors, including its distance to the host star and its mass.\par
It is only possible to estimate the planetary mass, as well as the separation of the planet to its host star from microlensing observations, if the Einstein radius and the distance of the lens to the observer are known. To estimate this distance, a method known as parallax is used. When observing an object from two different locations, it appears at different angular positions with respect to the background. Using simple triangulation, the distance to this object can then be calculated \citep{Yee2015CriteriaSurveys}.\par
Traditionally the parallax, and hence the distance to a star, is calculated by taking two measurements six months apart \citep{Yee2015CriteriaSurveys}. However, the duration of microlensing events can be approximately between several days and one month, meaning only a very small parallax is achieved and underconstrained results are gathered.\par
This method has still been rather successful as the ongoing Optical Gravitational Lensing Experiment (OGLE), in operation since 1992, shows. This experiment is an Earth-based project using microlensing which is responsible for the discovery of numerous exoplanets ranging from low mass planets around low mass stars to giant planets \citep{Kondo2021OGLE-2018-BLG-1185bDwarf,Blackman2021OGLE-2017-BLG-1434Lb:Optics,Gaudi2006ProspectsEffect}.\par
However, space-based observatories would be very beneficial to augment parallax measurements, as discussed in \cite{Bachelet2018MeasuringObservatories}. Arcanum could be used as such a space observatory, capable of calculating parallaxes by operating microlensing observation proposals after its primary goals are complete. As Arcanum will be located in an orbit around Neptune where the influence of the zodiacal light is negligible, reduced by a factor of 100 at Saturn alone, and as it will operate at a distance from the Sun of much greater than 1AU, the magnitude of the parallax generated and the clarity of any measurements should be sufficient to ensure reliable results.\par
James Bock, concept leader of the ZEBRA (Zodiacal dust, Extragalactic Background and Reionization Apparatus) instrument study \citep{Bock2012AstronomicalSystem} proposed that after the primary objective of measuring the Extragalactic Background Light (EBL) is achieved, the parallax between Saturn and Earth can be used for microlensing observations.\par
It could be expected that the Arcanum spacecraft could perform similar observations with an increased resolution when compared to the ZEBRA instrument, therefore further supporting the \lq{}multirole\rq{} designation of the spacecraft. Previous attempts to investigate such instruments, and the funding of such studies by NASA, acts as a reassuring sign for the legitimacy of this technique for exoplanet observation, and presents a natural and appropriate ability for Arcanum to possess.
\subsection{Verification of exoplanet detection techniques}
A potential goal for the telescope is to verify techniques used in exoplanet detection, taking advantage of its 30AU distance to Earth. It would benefit exoplanet scientists well if the Earth could be studied from afar to verify spectroscopic techniques and \lq{}wobbling\rq{} detection. For example, Earth’s atmosphere could be analysed using Arcanum’s spectrographs and compared with known values to determine the instrument and technique accuracy, for calibration and method refinement.
\section{Selection of Instruments}
\subsection{Seismometers}
The primary instrument of the penetrator probes is the seismometers. These devices will allow the subsurface of Triton to be analysed to determine whether Triton is still a volcanically active body and to further investigate its formation, interior structure and composition. There is a large amount of ongoing discussion as to their uses in determining the origin and history of planetary objects in general, as well as continuous monitoring to observe ongoing seismic activity, Earth-based and otherwise \citep{Erkan2008AMethods,Lorenz2011PlanetaryFuture}.\par
There are several types of subsurface imaging techniques that are commonly used:
\begin{itemize}
    \item Ground Penetrating Radar on the surface (GPR)
    \item Remote Sensing via satellites
    \item Magnetotellurics
    \item Electrical resistivity tomography (ERT)
    \item Seismometers
\end{itemize}
In comparison, the seismometers are the more desired method of data collection and are ideal for Triton when compared to the other instruments. GPR has its own advantages including non-invasive analysis of up to 30m beneath the surface. This can be seen with the Mars 2020 rover’s RIMFAX instrument, the first use of radar on the surface of Mars to analyse the structural features beneath the surface as well as determine the presence of resources such as water \citep{Hamran2020RadarExperimentRIMFAX}. Another example is China’s Chang’E-4 (CE-4) in 2019, which landed on the far side of the Moon and successfully used Penetrating Radar on the Yutu-2 Rover to image internal structure and thickness \citep{Li2020TheRadar}.\par
Seismometers, Accelerometers and Geophones all use inertial reference mechanics to record data but have different sensitivities and functions. The seismometers are the most sensitive option and provide a more defined image of the subsurface layers and seismic activity \citep{Hutt2010GuidelinesAccelerometers,Hou2021MEMSSeismometers}. All instruments will be required to have a high shock tolerance when launched from the lander due to the high speed of impact \citep{Hopf2010ShockApproach}.\par
Key seismometer data collection techniques would include:
\begin{itemize}
    \item Ambient Noise Tomography via large network array of seismometers \citep{Hons2008SeismicData,Ritzwoller2011AmbientArray}
    \item Seismic Attenuation Profile \citep{Tsuru2017GeophysicalProfiling}
\end{itemize}
On Earth, geoscientists use seismic devices to analyse various geological environments around the globe for vital issues involving Earth Science. Over the past 60 years, seismology has played an important role in geophysical observations for the South Pole \citep{Anthony2021SixObservations}. Antarctic ice penetrators applied to the ice sheets supply glaciological data which can be used to survey the conditions of the glaciers in response to the issue of climate change \citep{MITHaystackObservatory2021SGIP:Observatory}. This can be achieved by determining characteristics such as the thickness of the ice, the water content and any internal structures such as buried crevasses \citep{Kanao2014SeismicityOcean}. The use of such apparatus in an icy, frozen environment is also convenient as a potential model for Triton’s environment.  Most recently in 2014, an array of 100 temporary seismic stations were deployed to analyse the seismic coverage and noise of Antarctica \citep{Anthony2014TheAntarctica}. Furthermore, the Antarctic Network of Unattended Broadband Integrated Seismometers (ANUBIS) and the Federation of Digital Seismograph Networks are prime examples of the uses of a large network of seismic probes and supports proposed networks of penetrators on the Lunar \citep{Nunn2020MoonShake:Penetrators} and Martian \citep{Smith2011Multi-SitePenetrators} surfaces. ANUBIS, which was created in 1998 and ran for several years, was especially beneficial in determining the structure and properties of the crust and mantle of a planetary body \citep{Anandakrishnan2000DeploymentAntarctica}. Additionally, seismic devices situated close to volcanic sites provide constant observations of volcanic activity such as the magma flow, surface deformation and tomographic mapping of volcanic conduits. Others are also used to monitor earthquakes, providing accurate data of the size and location of the seismic waves to give early warnings.\par
In summary, seismometers are essential for the mission to determine the subsurface features of Triton. Future discussions on the choices between larger penetrators or microprobes, as well as suitable sites for analysis and whether an array for analysis is suitable, are required. Technology is currently being developed for future Mars and Lunar missions which will allow these instruments to make use of improved analysis techniques, enhancing data received and processed whilst also becoming more capable of handling for high-velocity impacts into the surface.
\section{Conclusions}
The onset of a new era in spaceflight, prompted particularly by revolutions in the space launcher market, will enable an increased launch cadence of our heaviest science-centric uncrewed spacecraft. The Arcanum mission aims to highlight this relatively imminent and dramatic change by showcasing the now-feasible systems which can be sent to the furthest reaches of our Solar System, and the comprehensive instrument suites they will be capable of carrying. A distributed surface science experiment at Triton has been shown to be not only possible, but also highly attractive, giving unparalleled insight into what is thought to be a captured KBO of astrobiological, geological and astrodynamical interest. The design for such an experiment, including component level spacecraft design and deployment can be demonstrated to be realistic, and the further justification of a large, predominantly optical, telescope in the outer Solar System continues to complete the mission proposal of Arcanum. Future work will make public simulations of the transfer trajectory, delta-V requirements for the spacecraft, and detailed numbers of the mass budget for each spacecraft.
\bibliographystyle{unsrtnat}
\bibliography{references}
\end{document}